\documentclass[aps,prl,twocolumn,showpacs,groupedaddress,floatfix]{revtex4}  % for review and submission 
\usepackage{graphicx}  % needed for figures
\usepackage{dcolumn}   % needed for some tables
\usepackage{bm}        % for math
\usepackage{amssymb}   % for math

\begin{document}

% The following information is for internal review, please remove them for submission
%\leftline{Version 06 as of \today} 
%\leftline{Primary authors: Tae Jeong Kim, Suyong Choi}
%\leftline{To be submitted to PRL}
%\rightline{Comment to {\tt d0-run2eb-005@fnal.gov}}
%\rightline{by January 17th, 2008}

% the following line is for submission, including submission to the arXiv!!
\hspace{5.2in} \mbox{Fermilab-Pub-08/058-E}

\title{
Search for pair production of doubly-charged Higgs bosons

in the $H^{++}H^{--} \to \mu^+\mu^+\mu^-\mu^-$ final state at D0
}
% LIST_OF_AUTHORS_R2.TEX               2/19/08              
%
\author{V.M.~Abazov$^{36}$}
\author{B.~Abbott$^{75}$}
\author{M.~Abolins$^{65}$}
\author{B.S.~Acharya$^{29}$}
\author{M.~Adams$^{51}$}
\author{T.~Adams$^{49}$}
\author{E.~Aguilo$^{6}$}
\author{S.H.~Ahn$^{31}$}
\author{M.~Ahsan$^{59}$}
\author{G.D.~Alexeev$^{36}$}
% ...

\affiliation{\vspace{0.1 in}(The D\O\ Collaboration)\vspace{0.1 in}}
% ...
\affiliation{$^{6}$University of Alberta, Edmonton, Alberta, Canada,
                Simon Fraser University, Burnaby, British Columbia, Canada,
                York University, Toronto, Ontario, Canada, and
                McGill University, Montreal, Quebec, Canada}
% ...
\affiliation{$^{29}$Tata Institute of Fundamental Research, Mumbai, India}
% ...
\affiliation{$^{31}$Korea Detector Laboratory, Korea University, Seoul, Korea}
% ...
\affiliation{$^{36}$Joint Institute for Nuclear Research, Dubna, Russia}
% ...
\affiliation{$^{49}$Florida State University, Tallahassee, Florida 32306, USA}
% ...
\affiliation{$^{51}$University of Illinois at Chicago,
                Chicago, Illinois 60607, USA}
% ...
\affiliation{$^{59}$Kansas State University, Manhattan, Kansas 66506, USA}
% ...
\affiliation{$^{65}$Michigan State University,
                East Lansing, Michigan 48824, USA}
% ...
\affiliation{$^{75}$University of Oklahoma, Norman, Oklahoma 73019, USA}
% ...
  % input Dzero author list
%\input list_of_visitor_addresses_r2.tex

\date{March 11, 2008}

\begin{abstract}
We report the results of a search for pair production of
doubly-charged Higgs bosons via  
$p\bar{p} \to H^{++}H^{--}X \to \mu^{+}\mu^{+}\mu^{-}\mu^{-}X$
at $\sqrt{s} = 1.96$~TeV.
We use a dataset corresponding to an integrated 
luminosity of 1.1 fb$^{-1}$ collected from 2002 to 2006 
by the D0 detector at the Fermilab Tevatron Collider.
In the absence of an excess above the standard model background,
lower mass limits of \mbox{$M(H^{\pm\pm}_L) > $ 150~GeV/$c^2$} 
and  \mbox{$M(H^{\pm\pm}_R) > $ 127~GeV/$c^2$}
at 95\% C.L. are set, respectively, for left-handed and right-handed
doubly-charged Higgs bosons assuming a 100\%  branching ratio into muons.
\end{abstract}

\maketitle 

%\section{\label{sec:level1}First-level heading}
% sections are not used for PRL papers
In the standard model (SM) of electroweak interactions, 
elementary fermions and bosons acquire mass via a weak isospin scalar doublet.
This mechanism results in the existence of an additional particle, the Higgs boson,
which has not yet been observed. Extensions of the Higgs sector involving
higher isospin multiplets
predict the existence of doubly-charged Higgs bosons which can be relatively light and hence accessible at current 
experimental facilities. 
Doubly-charged Higgs bosons appear in many scenarios
such as left-right symmetric models~\cite{left},
Higgs triplet models~\cite{triple}, and Little Higgs models~\cite{little}.
At the Fermilab Tevatron Collider, the two main production mechanisms are
pair production via
$p\bar{p} \to Z/\gamma^{*}X \to H^{++}H^{--}X$
and single production via $WW$ fusion,
$p\bar{p} \to W^{\pm}W^{\pm}X \to H^{\pm\pm}X$.
However, higher isospin Higgs multiplets
are generally severely constrained by \mbox{$\rho \equiv m^2_W / (\cos\theta_{W}m_Z)^2 = 1$} at tree level.
The existing phenomenological and theoretical constraints
are easily satisfied   
when the $W^{\pm}W^{\pm} \to H^{\pm\pm}$ coupling is vanishing~\cite{hunter}.
If the $H^{++}$ coupling to $W$ boson pairs is suppressed,
the dominant final states are expected to be like-sign lepton pairs.
Left-handed $(H^{\pm\pm}_L)$ and right-handed $(H^{\pm\pm}_R)$ states are distinguished by their
coupling to left-handed and right-handed leptons, respectively.
The pair production cross section for
left-handed doubly-charged Higgs bosons for \mbox{$100 \leq M(H^{\pm\pm}) \leq 200$~GeV/$c^2$}
is about a factor two larger than that for the right-handed states
due to different couplings to the intermediate $Z$ boson~\cite{nlo}.
Previous searches for $H^{\pm\pm}$ have been performed by the LEP collaborations~\cite{lep} 
in $e^+e^-$ collisions and by the D0~\cite{d0} and CDF~\cite{cdf} collaborations at the 
Tevatron $p\bar{p}$ collider.
This Letter presents the results of a direct search for 
$p\bar{p} \to H^{++}H^{--}X$ with $H^{\pm\pm} \to \mu^{\pm}\mu^{\pm}$
by the D0 collaboration with improved sensitivity.

The main D0 detector systems are a central tracking system, 
a liquid-argon and uranium calorimeter, and a muon detector~\cite{det}.
The central tracking system consists of the silicon microstrip tracker (SMT) and
the central fiber tracker (CFT) surrounded by a 2 T solenoidal magnet,
with designs optimized for tracking and vertexing capability
at pseudorapidity~\cite{eta}
$|\eta| < 3$ and $|\eta| < 2.5$, respectively.
The liquid-argon and uranium calorimeter has a central calorimeter (CC)
covering a region up to \mbox{$|\eta| \approx 1.1$} and
two end calorimeters (EC) extending
the coverage to $|\eta| \approx 4.2$, with each housed in a separate cryostat~\cite{cal}.
The muon detector has layers of proportional drift tubes and scintillation counters 
before and after a 1.8 T iron toroid~\cite{mudet}.
This analysis is based on the Run II data set
collected with the D0 detector at the Fermilab Tevatron Collider 
at $\sqrt{s} = 1.96$~TeV from  April 2002 to February 2006
corresponding to 1.1~fb$^{-1}$.
Events are collected using a suite of
dimuon and single muon triggers.

In the previous D0 analysis~\cite{d0},
two like-sign muons were required in the final state.
In this analysis, we require a third muon, which
increases the sensitivity by decreasing backgrounds.
We follow five steps to select events.
In the first step (S1),
events are required to have at least two muons.
Each muon must have a transverse momentum $p_T > 15$~GeV/$c$
and $|\eta| < 2.0$.
Muons are selected using patterns
of hits in the wire chambers and scintillators in the muon system.
Each muon must be matched to a track in the central tracker with at least five hits in the CFT layers and
at least two hits in the SMT layers.
Muons from cosmic rays are removed by using a timing information on the hits in the scintillator layers.

\begin{table*}
\caption{The expected numbers of events for a signal with $M(H_L^{\pm\pm}) = 140$~GeV/$c^2$ and
background and the number of observed events after each selection step.
The statistical and systematical uncertainties are combined in the table.
}
\begin{center}
\begin{tabular}{lr@{$\,\pm \,$}lr@{$\,\pm \,$}lr@{$\,\pm \,$}lr@{$\,\pm \,$}lr@{$\,\pm \,$}l}
\hline\hline
Selection & \multicolumn{2}{c}{Preselection} & \multicolumn{2}{c}{Isolation} & \multicolumn{2}{c}{$\Delta\phi$ $<$ 2.5} & \multicolumn{2}{c}{Like sign} & \multicolumn{2}{c}{Third muon} \\
                      &   \multicolumn{2}{c}{S1}   &    \multicolumn{2}{c}{S2}    &   \multicolumn{2}{c}{S3}   &     \multicolumn{2}{c}{S4}    &    \multicolumn{2}{c}{S5}  \\\hline
  $Z/\gamma^{*}\to\mu^+\mu^-$    &  69181&4642  & 58264&3910 &  4936&333 & 5.3&1.6 & \multicolumn{2}{c}{$<$ 0.01} \\
  Multijet                 &  4492&120 & 194&18 & 18&2 & 6.3&0.8  &  0.2&0.1 \\
  $Z/\gamma^{*}\to\tau^+\tau^-$  &  328&25 & 269&21 & 20&3 &  \multicolumn{2}{c}{$<$ 0.01} &  \multicolumn{2}{c}{$<$ 0.01}  \\
   $t\bar{t}$         &  38&3 & 20&1  & 14&1  & 0.03&0.01 &  \multicolumn{2}{c}{$<$ 0.01} \\
   $WW$               &  40&3 & 34&2  & 20&1 &  \multicolumn{2}{c}{$<$ 0.01} &  \multicolumn{2}{c}{$<$ 0.01}  \\
   $WZ$               &  19&1 & 16&1  & 11&1 & 2.95&0.20 & 1.62&0.11 \\
   $ZZ$               &  10&1 & 9&1   & 5&1  & 0.63&0.05 & 0.47&0.03 \\
   Total background   &  74108&4644 &58806&3910 & 5024&333 & 15.2&1.8 & 2.3&0.2 \\
 Signal               &  20.5&2.7 & 18.5&2.4  & 16.3&2.1 &  11.6&1.5 &10.1&1.3  \\
   Data               &  \multicolumn{2}{c}{72974}  & \multicolumn{2}{c}{58763}    & \multicolumn{2}{c}{4558}   &       \multicolumn{2}{c}{16}  &   \multicolumn{2}{c}{3}    \\\hline\hline
\end{tabular}
\end{center}
\label{table:sel}
\end{table*}

In the second step (S2), isolation criteria based on the calorimeter and tracking information
are applied
to remove the background from multijet production with muons originating from in-flight decay of
pions or kaons, or from semi-leptonic decays of $B$ or $D$ mesons.
The sum of the transverse energies of the calorimeter cells in an annulus of radius \mbox{$0.1 < {\cal R} < 0.4 $},
where \mbox{${\cal R} = \sqrt{(\Delta\phi)^2 + (\Delta\eta)^2}$} and $\phi$ is the azimuthal angle,
around the muon direction is required to be
less than $2.5~{\rm GeV}$.
A similar condition is defined for the scalar sum of the $p_T$ of all tracks, 
excluding the muon in a cone of
radius $ {\cal R} = 0.5$ centered around the muon,
which must be less than $2.5~{\rm GeV/c}$.

Selection S3 reduces the remaining \mbox{$Z\to\mu^+\mu^-$} and multijet backgrounds.
The azimuthal angle $\Delta \phi$ between at least one pair of muons is required to be less than 2.5 radians,
since the two muons from $Z$ boson decays are mostly back-to-back.
This requirement also rejects
a fraction of the multijet background with nearly back-to-back muons.

Selection S4 requires
at least two muons to be of like sign.
The final selection (S5) requires a third muon,
satisfying the S1 selection and the isolation selection criteria S2 
but without the minimum hit requirement on the
central track. 

The dominant background in this analysis arises from
electroweak processes where real high $p_T$ muons are created 
from $W$ or $Z$ boson decays as well as non-isolated muons originating from jets.    
The SM backgrounds and signal processes are generated
with {\sc pythia}~\cite{pythia} and normalized using the theoretical cross section.
The \mbox{$Z/\gamma^{*} \to \ell^+\ell^-$} cross section 
is calculated at next-to-next-to-leading order (NNLO)~\cite{Z}.
The $t\bar{t}$ cross section is calculated at NNLO~\cite{ttbar} and
the $WW$, $ZZ$ and $WZ$ cross sections are calculated
with {\sc mcfm}~\cite{mcfm} at next-to-leading order (NLO). 
All samples are processed through the D0 detector simulation based on {\sc geant}~\cite{geant} and
the same reconstruction software as for the data. 
The muon reconstruction and isolation efficiencies differ between Monte Carlo (MC) and data,
and these differences are corrected.
Trigger efficiency corrections are not applied to the MC sample.
Instead, the MC sample are normalized to the
data using the $Z$ boson mass peak at the selection level S2. 

Another important background comes from multijet production, mainly
$b\bar{b}$ events decaying semi-leptonically into muons that appear isolated.
The multijet background is derived from the data sample with non-isolated muons
obtained by inverting the isolation requirements 
for both muons after the selection S1.
The efficiency of the isolation requirement is assumed to be identical 
for multijet events with like-sign and opposite-sign muon pairs. It is also 
assumed that all like-sign events after subtracting SM backgrounds are multijet events.
The SM backgrounds are subtracted in the following samples used for the multijet
background determination.
The total number of multijet events before the isolation requirement ($4492\pm120$)
is then given by the number of non-isolated events for all charge combinations multiplied by the ratio
of the total number of events to the number of non-isolated events in the like-sign sample.
The number of multijet events after the isolation requirement ($194\pm18$) is obtained
by multiplying this number with the isolation efficiency ($4.3\pm0.5$)\%, given by the ratio
of isolated to all like-sign multijet events.

\begin{table}[b]
\caption{The numbers of observed and expected background events
after each selection criterion
with the like-sign requirement applied together with S1.
The statistical and systematical uncertainties are combined in the table.
}
\begin{center}
\begin{tabular}{lr@{$\,\pm \,$}lr@{$\,\pm \,$}lr@{$\pm$}l}
\hline
\hline
 Selection         &\multicolumn{2}{c}{Preselection}  & \multicolumn{2}{c}{Isolation}& \multicolumn{2}{c}{$\Delta\phi$ $<$ 2.5}  \\
 (Like-sign)        & \multicolumn{2}{c}{S1 \& S4}      & \multicolumn{2}{c}{S2}  & \multicolumn{2}{c}{S3}  \\\hline
$Z/\gamma^{*}\to\mu^+\mu^-$    & 84&24      & 42&12    &  5.3&1.6\\
 Multijet                & 1620&34   & 70&5    &  6.3&0.8\\
$Z/\gamma^{*}\to\tau^+\tau^-$  &  3.2&1.3 & 0.2&0.3 & \multicolumn{2}{c}{$<$ 0.01} \\
  $t\bar{t}$        &  6.6&0.5  &  \multicolumn{2}{c}{$<$ 0.1}      & \multicolumn{2}{c}{$<$ 0.1}  \\
   $WW$             & 0.08&0.02 & 0.04&0.01 & \multicolumn{2}{c}{$<$ 0.01}   \\
   $WZ$             & 5.14&0.35 & 4.25&0.29 & 2.95&0.20 \\
   $ZZ$             & 1.12&0.08 & 0.90&0.06 & 0.63&0.05 \\
   Total background & 1720&41   &   117&13   &   15.2&1.8   \\
   Data             & \multicolumn{2}{c}{1678} &  \multicolumn{2}{c}{96} & \multicolumn{2}{c}{16}   \\\hline\hline
\end{tabular}
\end{center}
\label{table:sel_like}
\end{table}

A second instrumental background arises from $Z/\gamma^{*}\to\mu^+\mu^-$ events
in which the charge of one of the muons is misidentified.
The first source of charge misidentification is due to fewer CFT layers
at large $\eta$ and a consequent increase in the charge misidentification probability.
The second source affects very high $p_T$ tracks for which
the uncertainty on the measured curvature can cause charge misidentification.
The charge misidentification rate is obtained by dividing
the number of like-sign events (S1, S2 and S4) by
the number of events without the like-sign requirement (S1 and S2) 
in the dimuon invariant mass region above 70~GeV/$c^2$,
after subtracting the SM sources of background except $Z/\gamma^{*}\to\mu^+\mu^-$ events from the data.
This mass requirement removes most multijet background events in the low mass range.
From these ratios, 
we determine the average probability for charge misidentification in data and MC to be
$P_{\rm data}=(6.2 \pm 1.1) \times 10^{-4}$ and
$P_{\rm MC}=(3.1 \pm 0.4) \times 10^{-4}$, respectively,
assuming the multijet background is negligible.
The uncertainties are statistical.
Since the charge misidentification rate in MC is underestimated,
the ratio of $P_{\rm data}$ to $ P_{\rm MC}$
is taken as a correction equal to $2.0\pm0.4$.
This ratio is applied to the $Z/\gamma^{*}\to\mu^+\mu^-$ MC sample when estimating the like-sign contribution.

\begin{figure}
\begin{tabular}{cc}
\includegraphics[width=4cm,height=3.7cm]{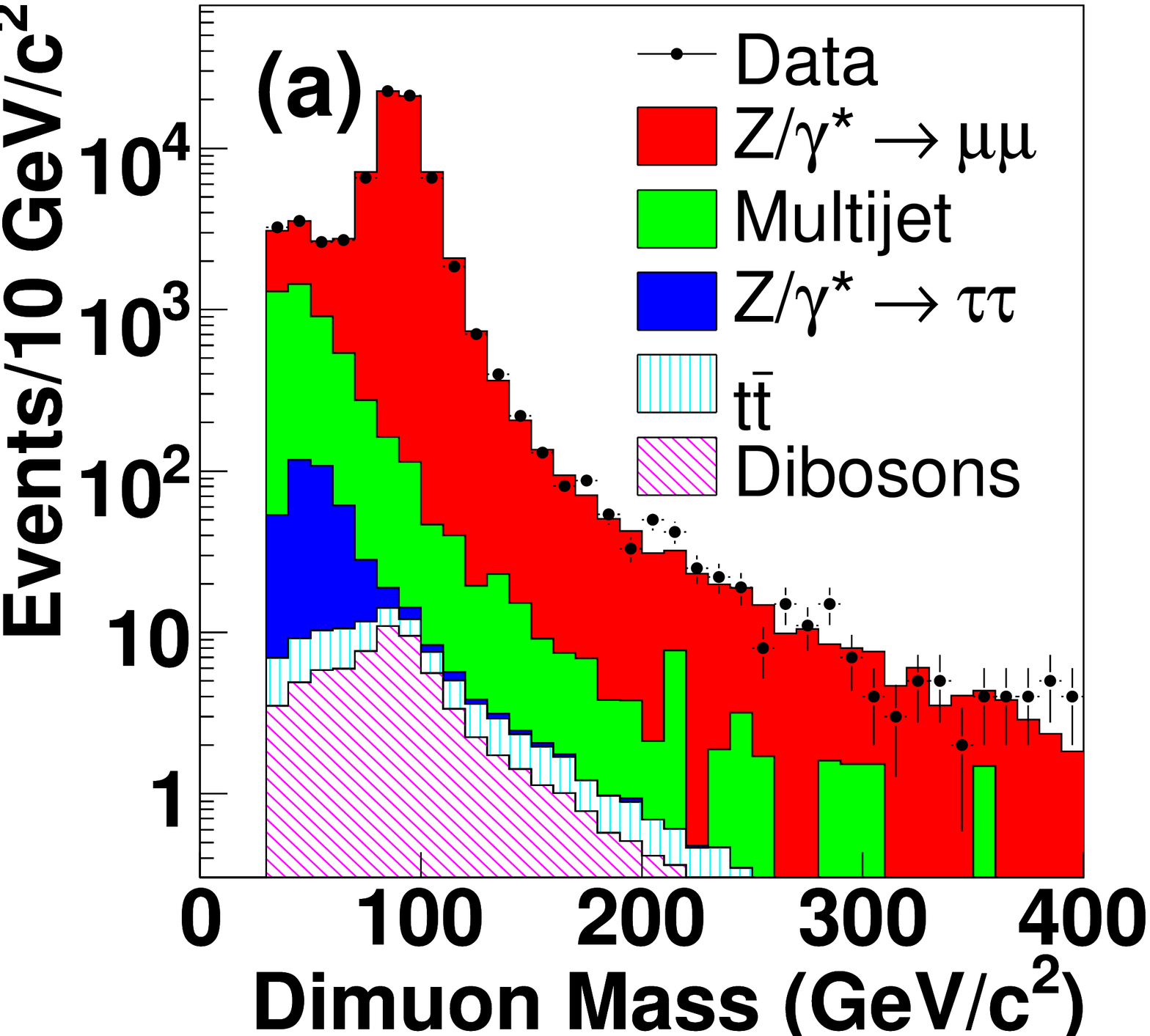} &
\includegraphics[width=4cm,height=3.7cm]{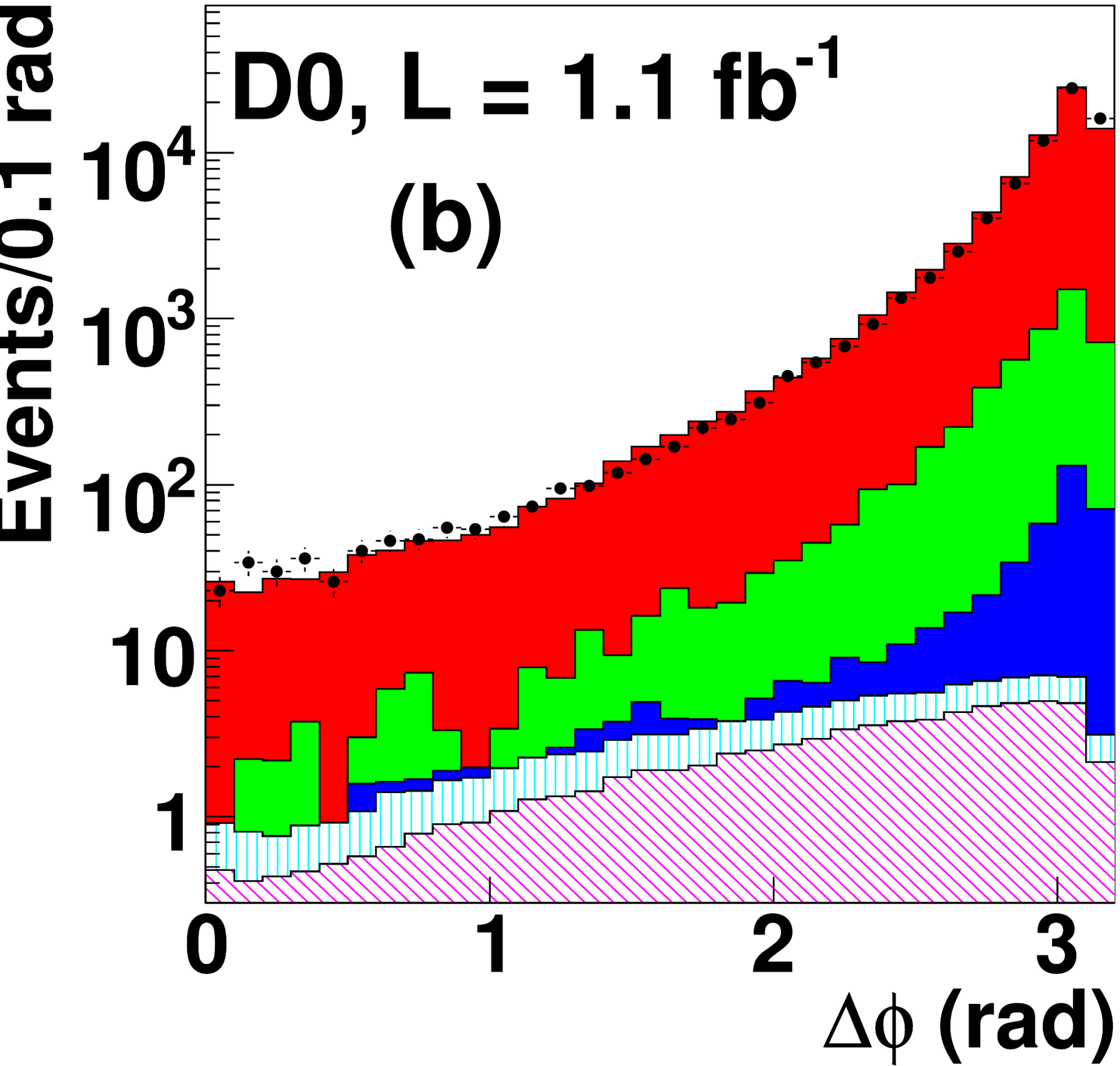} \\
\includegraphics[width=4cm,height=3.7cm]{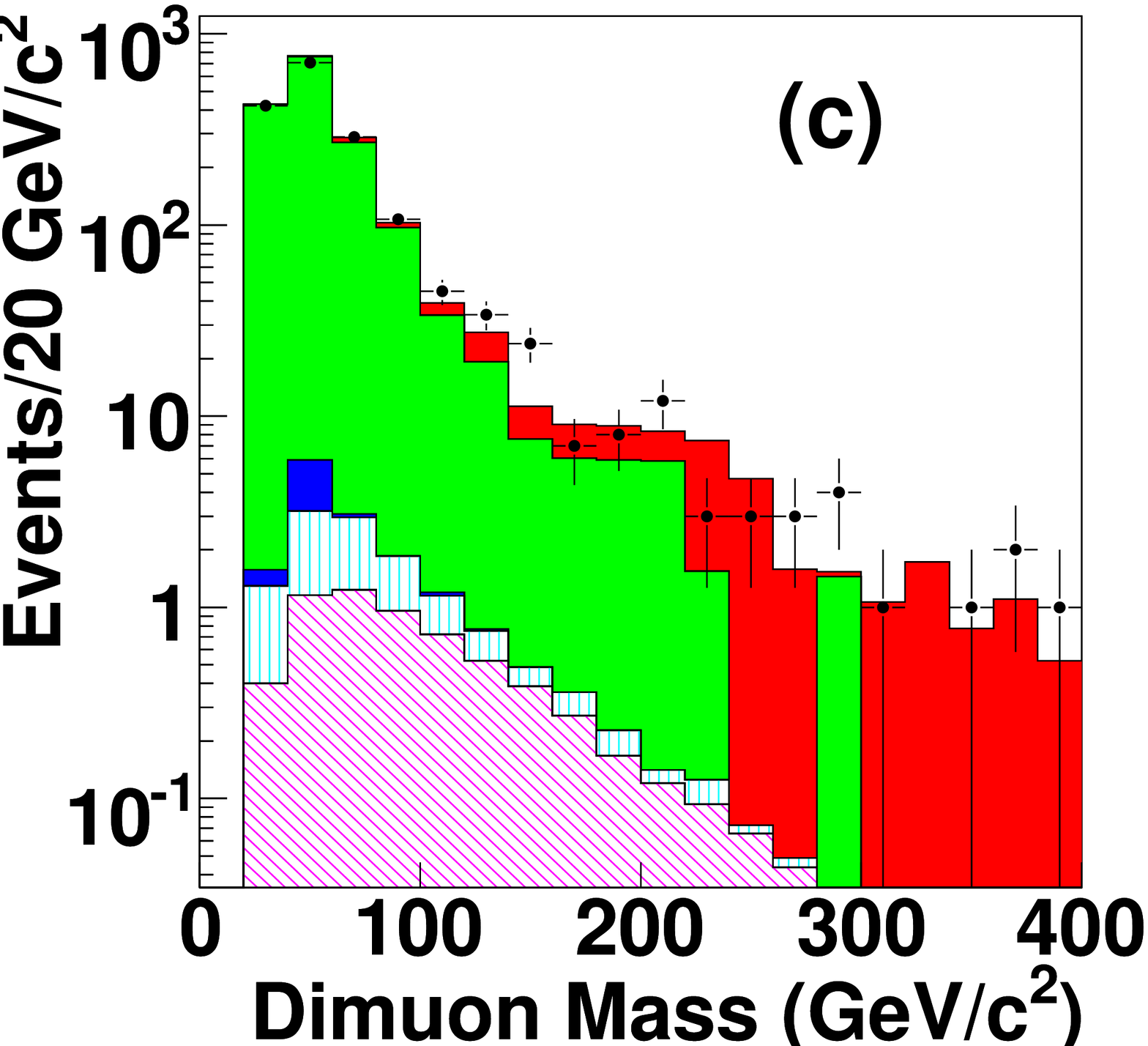} & 
\includegraphics[width=4cm,height=3.7cm]{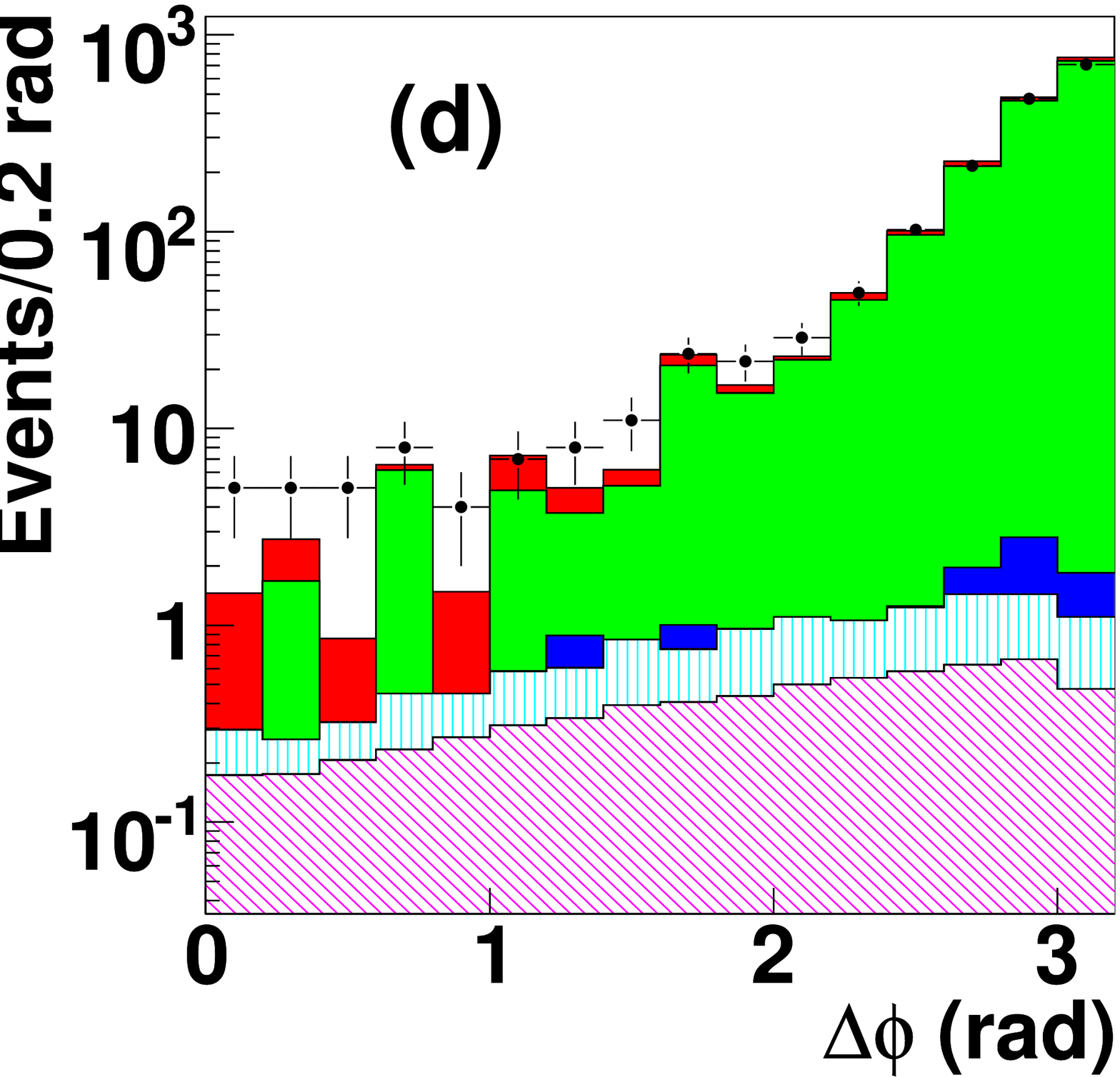} \\
\includegraphics[width=4cm,height=3.7cm]{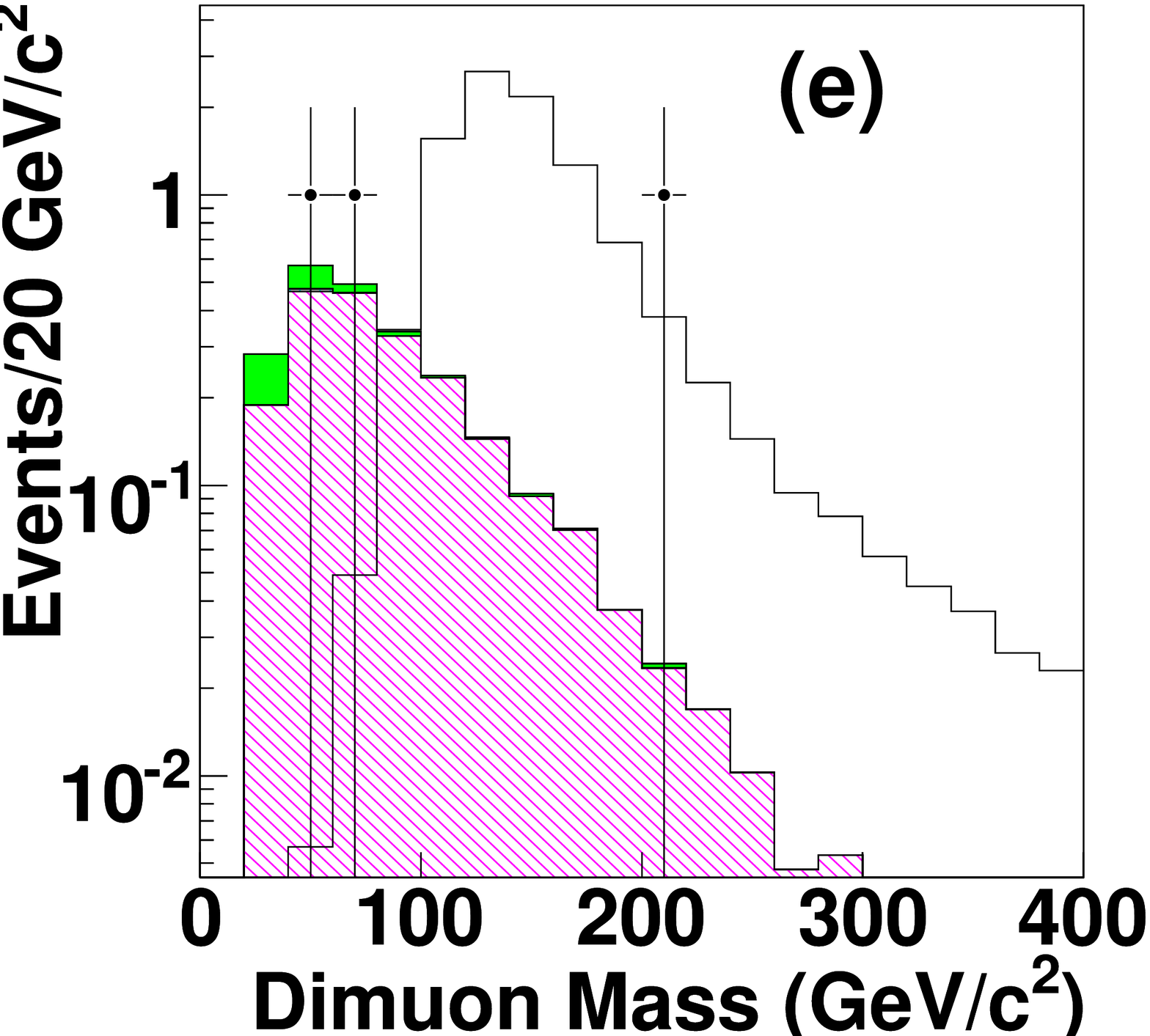} &
\includegraphics[width=4cm,height=3.7cm]{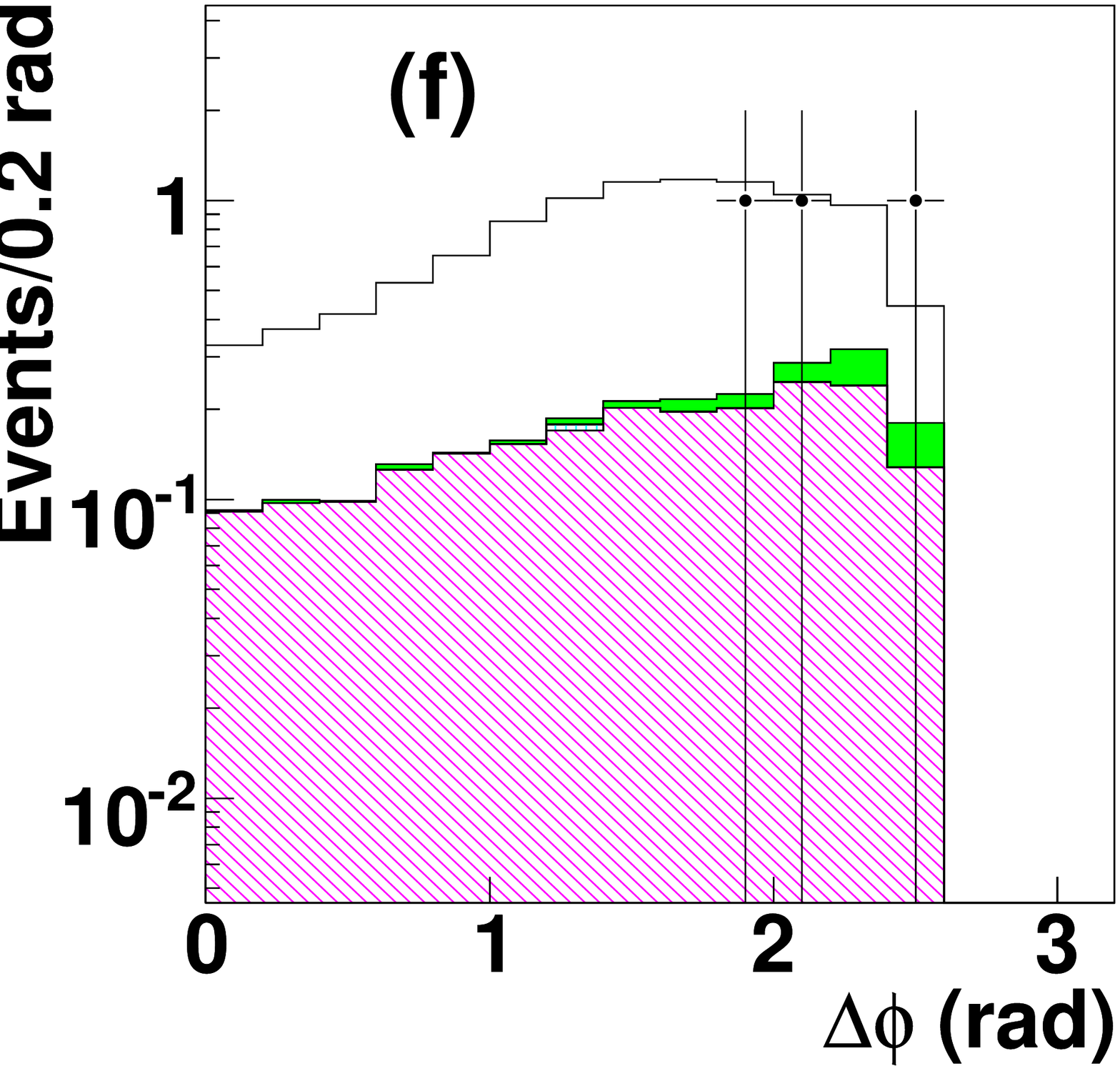} \\
\end{tabular}
\caption{\label{fig:num}Distributions of the dimuon invariant mass and $\Delta\phi$ between the two muons
for data compared to the sum of MC backgrounds after the selection S1 (a, b), 
the preselection S1 with the like-sign requirement S4 (c, d) and the final requirement S1--S5 (e, f).
The signal expected for a left-handed $H^{\pm\pm}$, with $M(H^{\pm\pm}) = 140$~GeV/$c^2$,
is also shown by the open histogram (e, f).}
\end{figure}

The distributions of dimuon invariant mass and $\Delta\phi$ 
after the selection S1 are shown in Fig.~\ref{fig:num} (a) and (b). 
The data are compared with the sum of the background contributions. 
For those events with more than one pair of muons fulfilling the selection criteria, 
the dimuon invariant mass and $\Delta\phi$ are calculated only for the pair with the highest individual momenta.
The numbers of remaining events after each selection are shown in Table~\ref{table:sel}. 
There is good agreement between data and the sum of the backgrounds.
Figure~\ref{fig:num} (c) and (d) show the dimuon invariant mass 
and $\Delta\phi$ distributions after the S1 and S4 requirements.
The excess of events at 150 GeV/$c^{2}$ has a significance of less than $2.6 \sigma$.
Table~\ref{table:sel_like} gives the individual like-sign backgrounds after the various selection stages.
This demonstrates that the like-sign backgrounds are well understood.

After all five selection criteria, three data events remain, in good agreement with the SM
background expectation of $2.3\pm 0.2$ events.
Total signal efficiencies are
\mbox{32\%--34\%} and are nearly independent of mass.
The dimuon invariant mass and $\Delta\phi$ distributions for these events are compared
to the sum of the backgrounds in Fig.~\ref{fig:num} (e) and (f).

Since no excess is observed, we use the dimuon invariant mass distribution in Fig.~\ref{fig:num} (e)
to compute upper limits on the production cross section times branching fraction
as a function of $M(H^{\pm\pm})$ using
the $CL_{S}$ method~\cite{CLs} as implemented in the {\sc mclimit} program~\cite{MCL}.
The expected rate for the signal as a function of $M(H^{\pm\pm})$ is
determined by the next-to-leading order (NLO) cross section~\cite{nlo} and
measured luminosity, corrected for the signal efficiency.

A number of systematic uncertainties on signal and background are taken
into account in the limit calculation.
The uncertainties on the correction of the muon identification are 2\% and 6\%
for backgrounds and signal, respectively.
The uncertainty on the isolation efficiency for the multijet background is 12\%.
The 20\% uncertainty on the correction for charge misidentification is included.
The uncertainty on the luminosity for signal is estimated to be 6.1\%~\cite{lumierror}.
The uncertainty on the normalization using NNLO MC SM background production cross sections
is taken to be 5\%.
The PDF uncertainties on the cross section for backgrounds are taken to be 4\%~\cite{pdf}.

\begin{figure}
\includegraphics[scale=0.32]{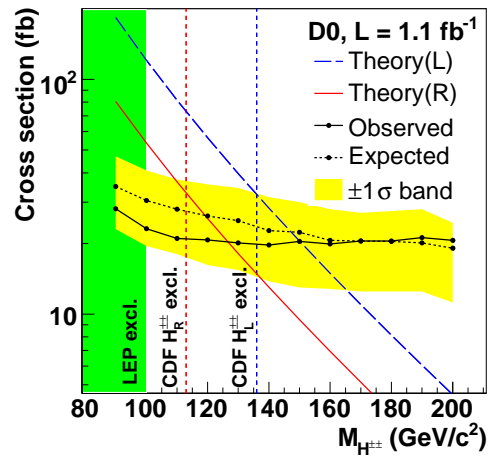}
\caption{\label{fig:final}The cross section limit as a function of the Higgs mass~$M(H^{\pm\pm})$
at the 95\% C.L.
The mass regions excluded by LEP and CDF are also shown.
The $\pm1 \sigma$ uncertainty on the expected limit is given by the yellow band (color online).}
\end{figure}

The cross section limit as a function of $M(H^{\pm\pm})$
is shown in Fig.~\ref{fig:final} together with the theoretical cross section for left- and right-handed 
doubly charged Higgs bosons. 
At the 95\% C.L., lower mass limits of 150~GeV/$c^2$
for left-handed and 127~GeV/$c^2$ for right-handed
doubly-charged Higgs bosons are obtained. This significantly extends the previous mass
limit~\cite{cdf} for a doubly-charged Higgs boson decaying into muons. 

% acknowledgement_paragraph_r2.tex                         2/19/08
%
We thank the staffs at Fermilab and collaborating institutions, 
and acknowledge support from the 
DOE and NSF (USA);
CEA and CNRS/IN2P3 (France);
FASI, Rosatom and RFBR (Russia);
CNPq, FAPERJ, FAPESP and FUNDUNESP (Brazil);
DAE and DST (India);
Colciencias (Colombia);
CONACyT (Mexico);
KRF and KOSEF (Korea);
CONICET and UBACyT (Argentina);
FOM (The Netherlands);
STFC (United Kingdom);
MSMT and GACR (Czech Republic);
CRC Program, CFI, NSERC and WestGrid Project (Canada);
BMBF and DFG (Germany);
SFI (Ireland);
The Swedish Research Council (Sweden);
CAS and CNSF (China);
and the
Alexander von Humboldt Foundation.


\begin{thebibliography}{99}
\bibitem{eta} 
The pseudorapidity is defined as 
$\eta = -\ln \left[ \tan \left( \frac{\theta}{2} \right) \right]$ as a function of the
polar angle $\theta$.
\end{thebibliography}
\end{document}